\documentclass[preprint,5p]{elsarticle}
\expandafter\let\csname equation*\endcsname\relax
\expandafter\let\csname endequation*\endcsname\relax
\usepackage{amsmath}
\graphicspath{{Figures/}}

\journal{Chaos, Solitons and Fractals}
\begin{document}

\begin{frontmatter}

\title{Transport in perturbed classical integrable systems: the pinned Toda chain}

\author[mymainaddress,mysecondaryaddress]{Pierfrancesco Di Cintio}
\corref{mycorrespondingauthor}
\address[mymainaddress]{Consiglio Nazionale delle Ricerche, Istituto di Fisica Applicata ``Nello Carrara'' via Madonna del piano 10, I-50019 Sesto Fiorentino, Italy}
\address[mysecondaryaddress]{Istituto Nazionale di Fisica Nucleare, Sezione di Firenze, via G. Sansone 1 I-50019, Sesto Fiorentino, Italy}
\cortext[mycorrespondingauthor]{Corresponding author}
\ead{p.dicintio@ifac.cnr.it}
\author[iub1,lepri1]{Stefano Iubini}
\address[iub1]{Dipartimento di Fisica e Astronomia, Universit\`a di Padova, via F. Marzolo 8 I-35131, Padova, Italy}
\author[lepri1,mysecondaryaddress]{Stefano Lepri}
\address[lepri1]{Consiglio Nazionale delle Ricerche, Istituto dei Sistemi Complessi, Via Madonna del Piano 10 I-50019 Sesto Fiorentino, Italy} 
\author[livi1,mysecondaryaddress,lepri1]{Roberto Livi}
\address[livi1]{ Dipartimento di Fisica e Astronomia and CSDC, Universit\`a di Firenze, via G. Sansone 1 I-50019, Sesto Fiorentino, Italy}
 \begin{abstract}
Nonequilibrium and thermal transport properties of the Toda chain, a prototype of classically integrable 
system, subject to additional (nonintegrable) terms are considered.  
In particular, we study via equilibrium and nonequilibrium simulations, 
the Toda lattice with a power-law pinning potential, recently 
analyzed by Lebowitz and Scaramazza [ArXiv:1801.07153]. We show that, according to general
expectations, even the case with quadratic pinning is genuinely non-integrable,
as demonstrated by computing the Lyapunov exponents, 
and displays normal (diffusive) conductivity for very long chains. 
However, the model has unexpected dynamical features and displays
strong finite-size effects and slow decay of correlations to be traced back to the 
propagation of soliton-like excitations, weakly affected by the harmonic pinning potential.
Some novel results on current correlations 
for the standard  integrable Toda model are also reported.
\end{abstract}
\end{frontmatter}
\section{Introduction}
Heat transport in classical low-dimensional lattices \cite{LLP03,DHARREV,Lepri2016} 
has been quite a debated problem in the last decades
and only recently unconventional effects (e.g., divergence of heat
conductivity, superdiffusion, the peculiar role played by nonlinearity, integrability and disorder,
etc.) have come to a satisfactory explanation. General theoretical approaches like 
nonlinear fluctuating hydrodynamics provides us
a way of predicting the basic features of anomalous heat conductivity in one-dimensional chains
of nonlinearly-coupled oscillators \cite{2012PhRvL.108r0601V,2014JSP...154.1191S,mendl2013,Cividini2017}, while rigorous predictions of the anomalous
behavior in stochastic conservative evolution of similar chains have been obtained \cite{2009JSP...134.1097B,2010ArRMA.195..171B,Lepri2009,Basile2016}.
Anyway, a good deal of numerical studies have paved the path of most of these achievements
and still allow us to obtain inference about many still open problems, like the way anomalous
behaviors depend on the kind of nonlinearity \cite{2005PhRvE..72c1202L,2008JSP...132....1L,2015PhRvE..91c2102L}, dimension (e.g. 1 or 2D) \cite{2017PhRvE..95d3203D}, disorder and the kind of the interaction (e.g.,  short-
or long-range; deterministic or stochastic) \cite{2018PhRvE..97c2102I,2015PhRvE..92f2108D}.

Although the general scenario for nonintegrable systems in low-dimensions
seems rather well established, 
there is a renewed interest in integrable systems, both classical and quantum.
The most famous classical example is the celebrated Toda model \cite{toda2012}.
The natural expectation would be to have ballistic 
heat transport  mediated by solitons \cite{toda2012}, as 
confirmed by the Mazur-bound type of arguments \cite{Mazur1969,Zotos02}.
Beyond this, there are intriguing features
that make this class of systems less obvious and worth being investigated.
For instance, the Hard Point Chain model (a gas of point particles colliding elastically
in one dimension) turned out to display anomalous conductivity even in the 
integrable limit of equal masses \cite{Politi2011}. Moreover, diffusive and 
and even anomalous 
spin transport was demonstrated for a classically integrable 
Landau-Lifshitz spin chain \cite{Prosen2013}.
On more general grounds, an interesting distinction between interacting and non-interacting integrable systems was recently proposed \cite{Spohn2018}. 
It was argued that in the latter case irreversible processes
may occur, yielding nonvanishing Onsager coefficients. Motion of thermally activated
Toda solitons would fall in this latter class due to 
the phenomenon termed nondissipative soliton diffusion \cite{Theo1999}, i.e., the stochastic sequence of spatial shifts experienced by the soliton as it moves through the lattice and interacts with other excitations without momentum exchange. 
A more recent study \cite{Kundu2016} showed that 
dynamical scaling is ballistic but also that correlation of hydrodynamic modes 
are non trivial.

The next question, which has an obvious physical relevance, concerns the role of 
perturbations away from integrability. It may be envisaged that strong enough 
changes of the Hamiltonian
of integrable one-dimensional systems will generically bring them back within one of the 
known classes of transport behavior. For instance, the ``diatomic'' versions of both
the Toda \cite{H99} and Hard Point Chain \cite{Delfini07a} models both display anomalous 
conduction in the Kardar-Parisi-Zhang universality class. Also, momentum-conserving stochastic 
perturbation of the Toda model yield energy superdiffusion \cite{Iacobucci2010}.
The situation may however be more subtle in
the limit of very small perturbations: for instance the FPU model in the region of
the parameter space where it is close to the Toda chain 
\cite{Benettin2013} displays an apparent diffusive 
transport \cite{2012PhRvE..85f0102Z,Das2014}. A similar feature occurs 
in the diatomic Hard Point gas with mass ratio close to one \cite{Chen2014}.

For all the above reasons, we are motivated to study in detail the 
classical Toda chain and its perturbations. The present work is organized as 
follows. In Section \ref{sec:model} we present the main model, the 
pinned Toda chain first considered in this context in \cite{2018arXiv180107153L}
and discuss qualitatively its dynamics. In Section \ref{sec:nonint},
we compute the Lyapunov exponents of the model to show that the pinning
generically destroys integrability. Proceeding in our analysis, we consider current 
correlations of the Toda model in Section \ref{sec:todacur}, while in in Section \ref{sec:pincur}
we study the effect of pinning and perturbations on these observables.
Thermal transport of the pinned Toda chain
is studied in Section \ref{sec:noneq} in the classical nonequilibrium setup, where the system
is in contact with two boundary reservoirs at different temperatures. 
Finally, a discussion  is given in the concluding Section. 

\section{The pinned Toda chain}
\label{sec:model}
In this paper we focus on the study of a one-dimensional Toda lattice of $N$ particles in the presence of a  pinning
potential, whose Hamiltonian reads \cite{2018arXiv180107153L}
\begin{equation}
\label{eq:hamilt}
H = \sum_{i=1}^N H_i=\sum_{i=1}^N \Big[ \frac{p_i^2}{2m} \,+\, V(q_{i+1} - q_i)\,+\,\frac{\nu^2}{z} |q_i|^z \Big],
\end{equation}
where $q_i$ and $p_i$ are canonically conjugated variables, $m$ is the mass of each oscillator, $z>1$ is a real exponent 
and $\nu$ a real frequency, and
\begin{equation}
\label{eq:todapot}
V(x) = \frac{a}{b}\, \left(e^{-bx}\,+\,bx-\,1\right)\,.
\end{equation}
We will set $a=b=1$ and $m=1$ and denote the energy density (energy per particle) 
as $\mathcal{E}=H/N$ henceforth. Most of the simulations are performed 
with periodic boundary conditions, unless otherwise stated. 

In the case of the standard Toda model ($\nu=0$), transport is ballistic,	 meaning that the 
finite-size thermal conductivity diverges linearly with the systems size $N$.
In linear-response language, this correspond to the existence of a nonvanishing
Drude weight, namely a finite value of the currents power spectra at zero frequency \cite{Mazur1969}.
This is roughly related to the density of thermally-excited solitons and 
implies a non-zero value of the current-current correlations at large times,
a generic feature of integrability \cite{Zotos02,Shastry2010,Spohn2018}. 

The presence of the pinning potential 
is expected to destroy the integrability of the Toda chain, while leaving 
only one conserved quantity: the total energy (\ref{eq:hamilt}).
 In particular, it
manifestly breaks space-translation invariance
and, accordingly, the conservation of total momentum, yielding an optical
dispersion in the harmonic limit, with a gap at zero wavenumber proportional to the
parameter $\nu$. 
From the point of view of transport, we thus expect that the 
model will display standard diffusive behavior, as known for other similar pinned chains, 
like the discrete Klein-Gordon and others \cite{LLP03,DHARREV}. 
Actually, the case $z=2$ is somehow special as for 
periodic boundary conditions it admits a second integral of 
motion, termed ``center of mass'' $h_c$ in \cite{2018arXiv180107153L} (see also Refs. \cite{lblb99,2004JSP...117..199L,2007PhRvE..75a1125P})
\begin{equation}
h_c= \frac{1}{2} \left(\sum_i p_i \right)^2 + \frac{\nu^2}{2} \left(\sum_i q_i \right)^2 .
\label{conserv}
\end{equation}
Note also that $h_c$ is conserved for an arbitrary choice of the interparticle
potential $V(x)$, as it can be easily ascertained by computing its time derivative.
Nonetheless, even in this case, we may expect diffusive transport, as observed in 
other similar models admitting two conserved quantities like the rotor (XY)
chain \cite{Giardina99,Gendelman2000} or the Discrete Nonlinear
Schr\"odinger equation \cite{Iubini2012,Iubini2013a,Mendl2015}.

\begin{figure}
\hspace{-12mm}
\includegraphics[width=0.62\textwidth]{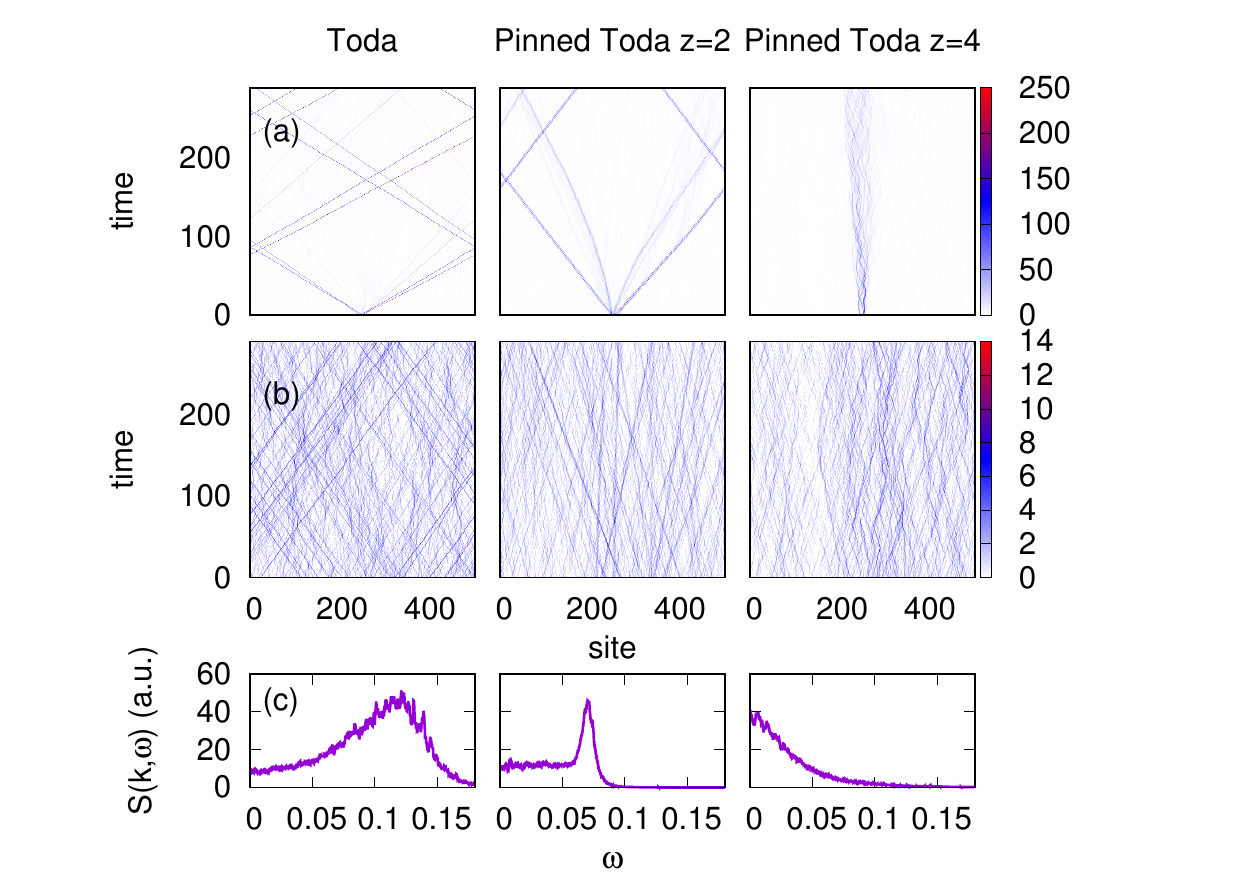}
\caption{Dynamics of the Toda $(\nu=0)$ and pinned Toda chains (with $\nu=1$) from microcanonical simulations
at energy density $\mathcal{E}=1$ for chains of length $N=512$ with periodic
boundary conditions; 
(a) and (b) are space-time plots of the local energy density.
In (a) the initial condition is generated by rescaling the momenta of the central 20
sites of a thermalized state by a factor 10 while (b) is a generic microcanonical state
obtained assigning random velocities to all the particles.
Lower panels (c) are the structure factors versus frequency $\omega$
for a fixed wave-number $k= \pi/256$. } 
\label{f:proff}
\end{figure}

The dynamics of the model is illustrated in Fig. \ref{f:proff}. In the upper panels 
we report the space-time evolution of the local energy field $H_i$ 
for the Toda chain, and the pinned Toda chain with $z=2$ and 4 and for two 
types of initial conditions, a  locally perturbed thermal state, Figs.\ref{f:proff}a,  
and an unperturbed one, Figs.\ref{f:proff}b (by thermal state we mean a
generic configuration obtained evolving the dynamics from random initial conditions).
In both simulations, the model with $z=2$ behaves somehow more similar to the
unperturbed Toda model than in the case $z=4$. 
Actually, a significant ballistic propagation is observed, 
akin to soliton motion, drastically
different from the diffusive-like spreading shown in the rightmost panels.
Also in the central panel of Fig. \ref{f:proff}a a clear
dispersion of velocities is present, a feature different from what seen
in generic nonintegrable chains, like the Fermi-Pasta-Ulam model.
This difference is distinctly captured by the dynamical structure factors 
(Figs.\ref{f:proff}c): $S(k,\omega)$ are defined as the modulus squared of the 
spatio-temporal Fourier transform $\hat q$ of the particle displacements $q_i(t)$
\begin{equation}\label{somega}
S({k},\omega)=\langle|\hat q({k},\omega)|^2\rangle.
\end{equation}
Since we are working with periodic boundary conditions, the allowed values of 
the wave number $k$ are always integer multiples of $2\pi/N$.
For $z=2$, $S({k},\omega)$ displays a narrow peak at finite frequency, whose position
is proportional to the wavenumber $k$. Clearly, this case is qualitatively more similar 
to the Toda case and definitely different from the chain with  $z=4$ that only
show a central diffusive peak.

The heat conduction problem for a pinned Toda chain coupled to external heat baths  has been 
studied in a recent paper by Lebowitz and Scaramazza \cite{2018arXiv180107153L}. 
The main surprising result there reported 
is that for $\nu\not=0$ and $z=2$ the model seems to exhibit ballistic transport
of energy, as it should be expected for an integrable model, like a chain of harmonic oscillators
or a pure Toda lattice. On the other hand, if $z=4$ the
ballistic transport disappears and one obtains temperature profiles that are closer
to the expected Fourier linear behavior (despite they still exhibit sensible deviations from that). Such numerical analysis, prompted the Authors to argue that the Toda chain with harmonic pinning might be integrable or, more likely, 
an example of a nonintegrable system without momentum conservation for which the heat flux is ballistic.
 
Altogether, this phenomenology and the results of \cite{2018arXiv180107153L},
suggest that transport in the harmonically pinned case 
may display some features of solitonic propagation, possibly related to its
underlying integrable model, that we investigate in the following.
\section{Nonintegrability of the pinned Toda chain}
\label{sec:nonint}
In this Section we first demonstrate that the pinned Toda chain is genuinely
nonintegrable also in the case $z=2$ through computation of Lyapunov spectra.
The spectra  $\lambda_i$, $i=1,\ldots2N$ are computed with the standard method \cite{Pikovsky2016}
for chains with
periodic boundary conditions and the same parameter 
values adopted in
\cite{2018arXiv180107153L}, as well as with different combinations of the energy density, particle number and frequency $\nu$.

The equations of motion $p_i = -\partial H / \partial q_i$ and $q_i = \partial H / \partial p_i$ and the associated tangent-space equations used to evaluate the Lyapunov exponents 
have been integrated with a standard 3-rd order symplectic algorithm \cite{1990CeMDA..50...59K,mclachlan1992accuracy} with fixed time step $\delta t =0.01$, while for the simulations presented in the following Sections we used 
the 4-th order scheme with $\delta t =0.05$.

\begin{figure}
\includegraphics[width=0.42\textwidth]{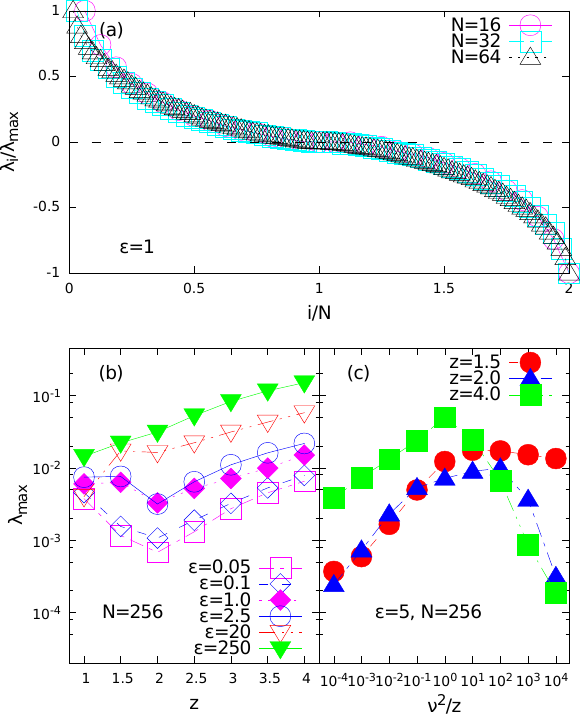}
\caption{(a) Lyapunov spectra (normalized to the maximum exponent $\lambda_{\rm max}$ and plotted versus the rescaled index $i/N$) of the Toda chain with harmonic pinning $(z=2)$ and 
$\mathcal{E}=1$ for $N=16,$ 32 and 64. For $N=256$:
(b) maximal Lyapunov exponent $\lambda_{\rm max}$ as function of the pinning potential exponent 
$z$, for different values of the energy density $\mathcal{E}=0.05,$ 0.1, 1, 2.5, 20 and 250, 
and fixed $\nu=1$; (c) $\lambda_{\rm max}$ as function of the ratio $\nu^2/z$ for $z=1.5$, 2 and 4
for fixed energy density $\mathcal{E}=5$.}
\label{fig2}
\end{figure}

In Fig. \ref{fig2} (a) we show the Lyapunov spectra of model (\ref{eq:hamilt}) with $z=2$ for
different values of $N$: they display the usual convergence to the limit form, typical of generic 
chaotic Hamiltonian system \cite{Livi1986}.
We also checked that in the harmonic pinning case there exist, beyond the two 
vanishing Lyapunov associated with energy conservation, an additional pair 
of zero exponents as it should in view of the existence of  
the additional conserved quantity $h_c$ given in Eq. (\ref{conserv}).

In addition, we have evaluated the dependence of 
maximum Lyapunov exponent, $\lambda_{\rm max}\equiv\lambda_1$ on the model parameters.  
In Fig. \ref{fig2} (b) we show $\lambda_{\rm max}$ as a function of $z$ for $N=256$ and $\nu=1$ for different values of the energy density in the range $5\times 10^{-2}\leq\mathcal{E}\leq 2.5\times 10^2$, 
while in Fig. \ref{fig2} (c) we show $\lambda_{\rm max}$ as a function of $\nu^2/z$ for $N=256$ and $\mathcal{E}=5$ for $z=1.5$, 2 and 4. 
In all cases $\lambda_{\rm max}$ is always positive, which further excludes the possibility that model described by the Hamiltonian (\ref{eq:hamilt}) can be integrable even for particular values of $z$. Remarkably, 
for low values of $\mathcal{E}$, $\lambda_{\rm max} (z)$ has an evident non-monotonic trend with a minimum in $z=2$, that gradually disappears for increasing values of $\mathcal{E}$. 
Such non-monotonicity with the energy density $\mathcal{E}$ is independent on the system
size $N$, but it is seemingly depending on the pinning strength $\nu$ for fixed values of $N$ and $\mathcal{E}$.  
This confirms that the harmonic pinning is somehow special from the dynamical point of 
view, as we will show in the following sections.
\section{Currents correlation: the Toda chain}
\label{sec:todacur}
We now turn to the main observable related to transport, namely current-current
correlations (or equivalently current power spectra) computed in equilibrium. Let us discuss first the case of the Toda chain. Although several studies
of dynamical correlations for various observable exist in the literature \cite{Schneider1980,Cuccoli1993,Theo1999}
it is useful to reconsider the behavior of flux correlations in detail.  

The Toda model has $N$ integrals of motion, and to each of them a local current 
can be associated through a continuity equation \cite{Spohn2018}. 
Usually one considers the total (volume-averaged) currents $J_k$ ($k=1,\ldots,N$)
that are directly related to the Onsager matrix. 
In principle one has to
deal with a $N\times N$ matrix of correlations $\langle J_k(t)J_l(0)\rangle$.

\begin{figure}
\includegraphics[width=0.45\textwidth]{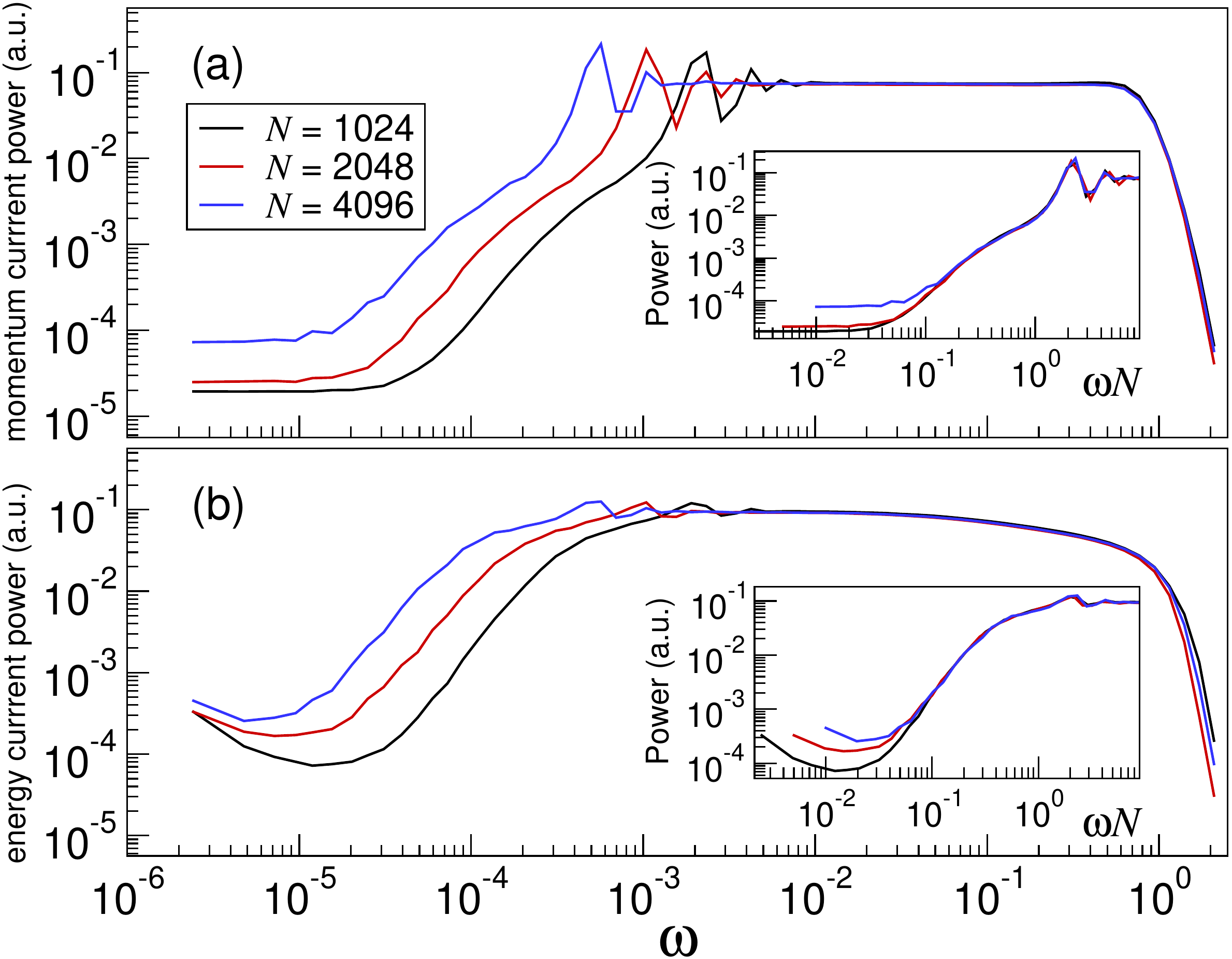}
\caption{Power spectra of the momentum- (a) and energy- (b) currents of the Toda lattice (see  
Eq. \ref{currents}) for different chain lengths $N$ and $\mathcal{E}=1$. Insets: same
data as a function of the rescaled frequencies $\omega N$.}
\label{fig3}
\end{figure}

Here we limit ourselves to considering the two simplest integrals, namely total momentum and
total energy, along with their respective total 
currents
\begin{equation}
J_1 = \sum_{i=1}^N F_i; \quad 
J_2=\frac{1}{2}\sum_{i=1}^N (\dot q_{i+1}+\dot q_{i})F_i.
\label{currents}
\end{equation}
where $F_i=-V'(q_{i+1}-q_{i})$ is the interparticle force.
In Fig. \ref{fig3} we report their power spectra  $\langle |	\hat J_{1,2}(\omega)|^2\rangle$
averaged over a long microcanonical trajectory started from a random Gaussian 
distribution of velocities (this may not be fully
representative of the real statistical ensemble, which for completely
integrable systems should be the Generalized Gibbs Ensemble \cite{vidmar2016generalized}).
Note that the value at $\omega=0$ (the Drude weight) is not 
vanishing but it is not reported in the figures.
From this analysis, there is an indication that both spectra vanish for $\omega \to 0$.
Actually, the typical frequency scale
below which the spectra approach a small value is of order $1/N$, as
demonstrated by 
the rescaled data in the insets of Fig. \ref{fig3}. 
It is natural to associate this time-scale to the typical time of transit of 
solitons across the chain.
If we take the $N\to \infty$ limit before the $\omega \to 0$ limit,
as we should in the calculation of transport coefficients via the Green-Kubo
formula, there would be a finite Onsager coefficient, yielding some entropy production on top
of the ballistic contribution. In this respect, the data are fully consistent with 
regarding the Toda chain as an interacting integrable model, as recently surmised in \cite{Spohn2018}.
\section{Currents correlation: the pinned Toda chain}
\label{sec:pincur}
\begin{figure}
\includegraphics[width=0.45\textwidth]{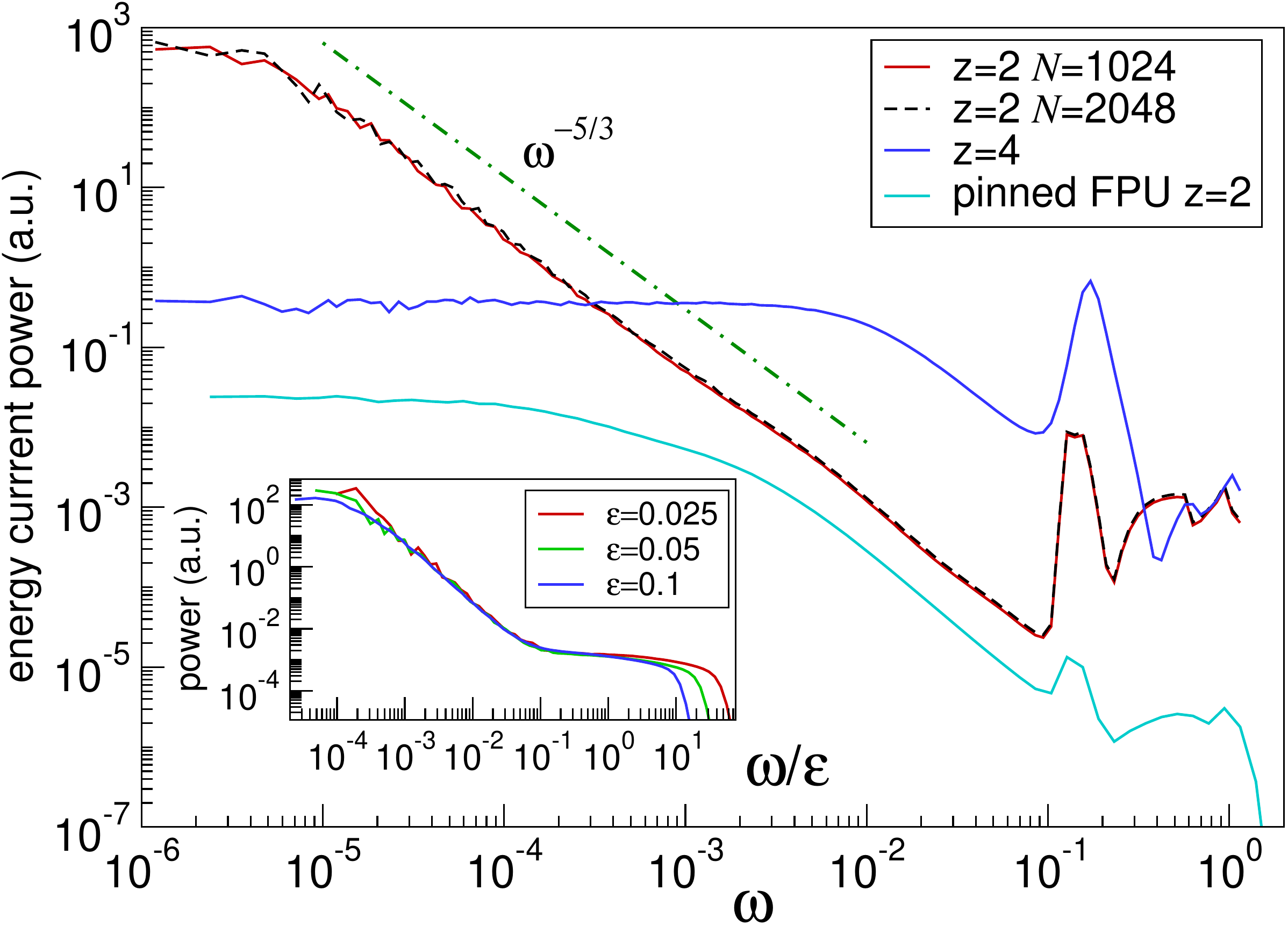}
\caption{Power spectra of the energy current for a pinned Toda lattice with $z=2$ and 4, $\mathcal{E}=1$ and $N=1024$ (solid lines) and $2048$ (dashed line).  The green dot-dashed line marks the power-law $\omega^{-5/3}$ trend of the spectra for $z=2$. For comparison,
the energy spectrum for a pinned FPU lattice with the same $\mathcal{E}$, $z=2$ and $N=1024$ (cyan line) is reported. Data has been shifted vertically for better readability.
Inset: power spectra of the energy current for a  Toda lattice 
perturbed by a momentum-conserving harmonic potential, Eq. (\ref{eq:hamiltper}).
The frequency axis is divided by the perturbation strength $\varepsilon$ for 
comparison.}
\label{fig4}
\end{figure}

Let us now turn to the effect of pinning. In this case we consider the energy current
correlation only. The results are shown in the main panel of Fig. \ref{fig4}.

For $z=4$, $\langle |\hat J_{2}(\omega)|^2\rangle$ approach a constant
for $\omega\to 0$ implying, as expected, a relatively  
fast decay to zero of correlations and finite thermal conductivity.

For the harmonic case $z=2$ the same conclusion holds, but the saturation
occurs below a much smaller value of frequency. In addition, and
somehow surprisingly, there exist a wide intermediate 
range of frequencies where the spectra display a power law 
\begin{equation}
\langle |\hat J_{2}(\omega)|^2\rangle \propto \omega^{-\theta};\quad \omega_1<\omega<\omega_2 
\label{power}
\end{equation}
with a nontrivial exponent $\theta\approx 5/3$. Such unusual behavior is very stable 
with the system size (the black dashed and the red solid curves, corresponding respectively to $N=1024$ and
$N=2048$, almost overlap) and it should 
be compared with the case $z=4$, where correlations decay much faster.
We also checked that simulations at lower energy density, e.g.  $\mathcal{E}=0.1$, 
yield similar results (data not reported).

To ascertain whether the underlying Toda potential is relevant we compared the
data with the FPU approximation of the Toda potential, namely by replacing
$V$ in Eq. (\ref{eq:hamilt}) with its power-series expansion truncated to fourth
order. Remarkably, the spectra change completely, yielding a faster
decay of correlations. From this we conclude that the presence of an underlying integrable
model is essential to observe the power-law behavior given by Eq. (\ref{power}).
The above results indicate that different classes of perturbations of the Toda
model may have drastically different and unexpected 
effect on transport. To illustrate this further let us for instance consider the Hamiltonian
\begin{equation}
\label{eq:hamiltper}
H_1 = \sum_{i=1}^N \Big[ \frac{p_i^2}{2m} \,+\, V(q_{i+1} - q_i)\,+\,
\frac{\varepsilon}{2} (q_{i+1} - q_i)^2 \Big] ,
\end{equation}
that has the usual three conservation laws (total energy, momentum and elongation) and is thus expected to 
belong to the Kardar-Parisi-Zhang universality class of nonintegrable 
one-dimensional models \cite{2014JSP...154.1191S}. 
In the inset of Fig. \ref{fig4} we show that this expectation 
is not verified, at least in the considered range, even for very small
strength of the perturbation $\varepsilon$. To compare the different simulations,
we rescaled the frequency axis by the perturbation strength $\varepsilon$. 
Empirically, we found a satisfactory data collapse that indicates that
the typical time-scale of decay is inversely proportional to $\varepsilon$.
We do not have an explanation for this observation. The main relevant 
point here is that the expected anomalous behavior is somehow not 
observed on the accessible time-scales in a way similar to other 
weakly perturbed integrable systems
\cite{Benettin2013,2012PhRvE..85f0102Z,Das2014,Chen2014}.

\section{Nonequilibrium properties}
\label{sec:noneq}
In this Section, we further demonstrate that transport in the harmonically pinned Toda chain
obeys Fourier's law, but that finite-size effects are indeed very relevant.
We have performed nonequilibrium numerical simulations  for system sizes much larger that
those studied in \cite{2018arXiv180107153L}, while integrating the equations of motion over definitely much longer  
time.  We have adopted free boundary
conditions for the oscillators at site 1 and $N$, while a fraction  $N_0=N/16$ of  oscillators at the
two chain ends have been independently thermalized by Maxwellian heat reservoirs at temperature $T_L$ and $T_R$, respectively, as in \cite{2018PhRvE..97c2102I}. 
\begin{figure}
\includegraphics[width=0.44\textwidth]{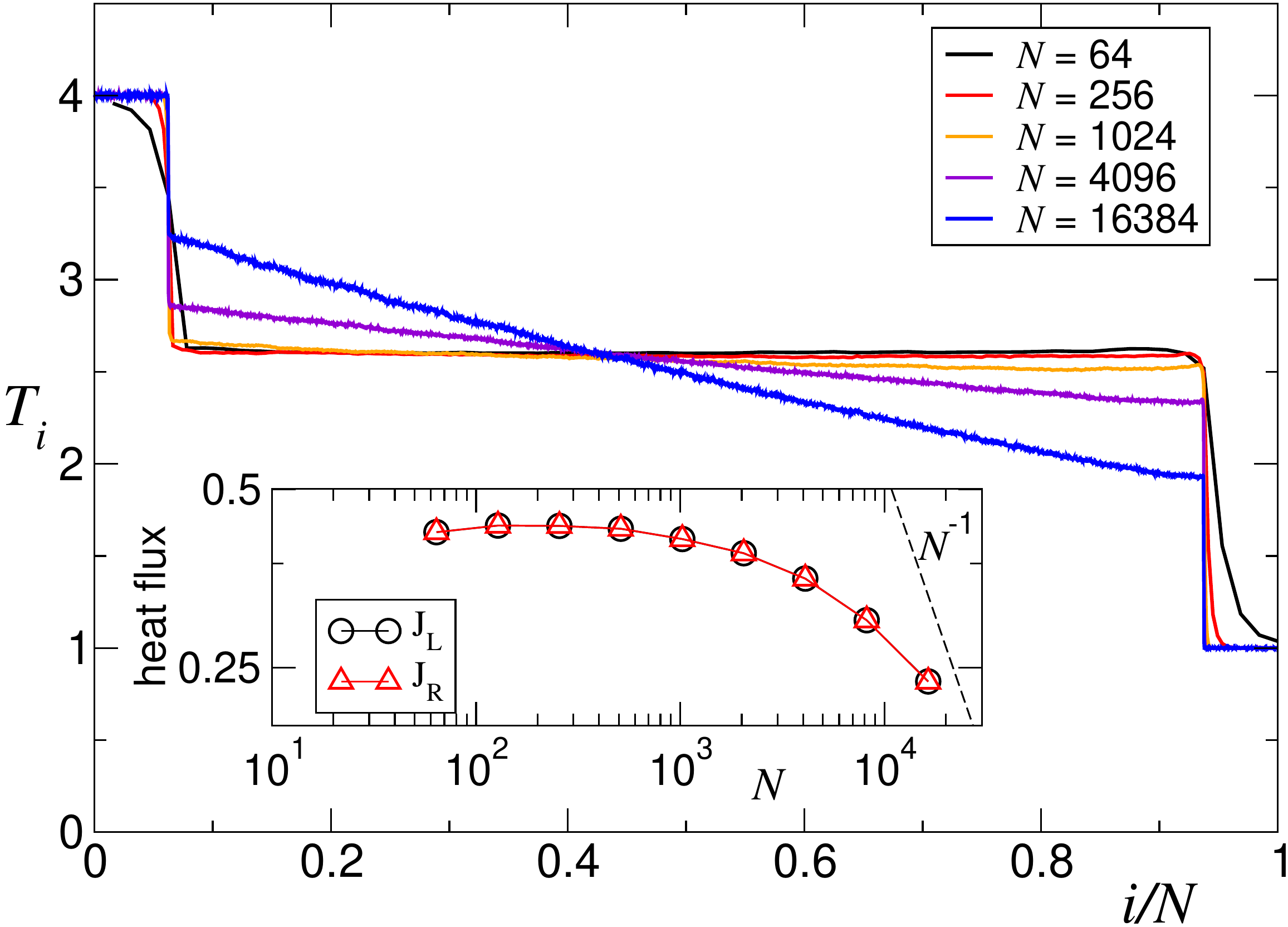}
\caption{Nonequilibrium simulation of the 
Toda chain with harmonic pinning $z=2$: Stationary temperature profiles (reported versus the rescaled variable $i/N$) as a function of the chain length $N$ for $T_L=4$ and $T_R=1$ (main plot). Left and right heat fluxes $J_{\rm left}$ and $J_{\rm right}$ as function of the system size $N$ (inset). In all cases the simulations refer to Fourier initial conditions, corresponding to a linear temperature profile that interpolates between $T_L$ and $T_R$, and are extended up to  $t_f=5\times 10^6$.}
\label{fig5}
\end{figure}

In Fig. \ref{fig5} we show the stationary temperature profiles
$T_i=\overline{p_i^2(t)}$, where the bar denotes the average over time $t$,
 for the case $z=2$, with $\nu=1$ and different chain sizes in the range $64\leq N\leq 16384$. In all cases the external temperatures were fixed to $T_L=4$ and $T_R=1$ and the total integration time $t_f$ was chosen equal to $5\times 10^6$ units, which was sufficient for reaching stationary conditions for all system sizes here considered (see below). 
 It can be seen clearly that even relatively long chains (up to $N\simeq 10^3$) exhibit almost flat temperature profiles. Conversely, only for $N > 10^4$ a linear temperature profile  interpolating between $T_L$ and $T_R$ is definitely distinguishable, although with significant finite-size effects in the form of temperature drops close to thermalized interfaces.
These jumps account for the well-known phenomenon of  an effective  impedance
between the heat reservoirs and the chain \cite{Aoki2001}, because of the mismatch of response
time between the reservoir time constant and the typical relaxation time scales of 
fluctuations, which is associated with the kind of nonlinear modes propagating
through the chain.  
 
 To complement our analysis, in the inset of Fig. \ref{fig5}  we present as function of $N$ the value of the left (circles) and right (triangles) heat fluxes $J_{L}$ and $J_{R}$, that were computed as the average energy exchanged by the system with each reservoir  per unit time \cite{2018PhRvE..97c2102I}. The two curves nicely overlap, thus confirming that the system has reached a nonequilibrium steady state. Consistently with the behavior of temperature profiles, the plateau initially visible for small $N$ (reminiscent of ballistic conduction of integrable systems \cite{LLP03}) starts to bend for $N> 10^3$, although with a trend quite far from the scaling expected
for standard diffusion, $J_{L,R}\sim N^{-1}$: see the black dashed line. 

\begin{figure}
\includegraphics[width=0.44\textwidth]{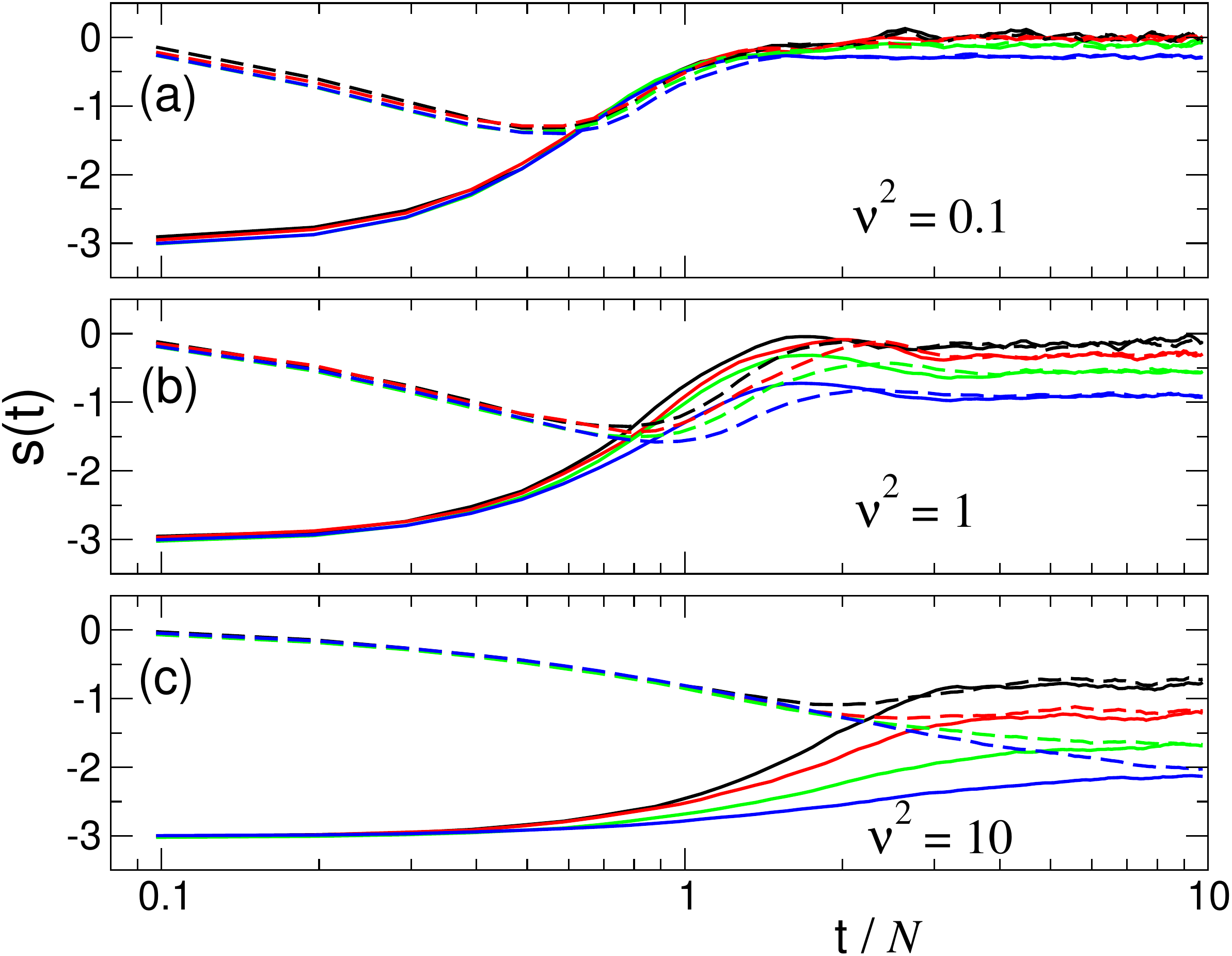}
\caption{Evolution of the slope $s(t)$ of temperature profiles for different models with $z=2$ starting from initial condition characterized by flat (dashed lines) and Fourier (solid lines) temperature profiles with $T_L=4$ and $T_R=1$ (see text).
Black, red, green and blue curves refer respectively to $N=1024,$ 2048, 4096 and 8192.
 For all curves, 
time has been rescaled to the number of sites $N$. Panels (a,b,c) refer to three 
increasing values of the pinning frequency $\nu$.}
\label{fig6}
\end{figure}

 We also addressed the question of how the system relaxes to the nonequilibrium
steady state and how the typical timescales depend on size and initial conditions.
In Fig. \ref{fig6} we report the slope $s(t)$ of the temperature profile in the
 bulk (excluding directly thermalized particles)
  as a function of the
rescaled integration time $t/N$ up to $t/N=10$. The two different sets of curves, dashed and solid, correspond to
different initial conditions, namely (i) a flat profile at $(T_R +T_L)/2$ and (ii) 
a Fourier profile,
linearly interpolating between $T_L$ and $T_R$, respectively. For a certain time $t$, $s(t)$ is obtained by fitting time-dependent temperature profiles $\tilde{T}(x,t)$ 
with linear functions $f(x,t)=s(t) x + c(t)$, where $c(t)$ is an offset and $x=(i-N_0)/(N-2N_0)$.
The function  $\tilde{T}(x,t)$ is computed 
by averaging local square momenta over temporal windows of length $0.1N$ and over an ensemble of 64 independent realizations 
of the dynamics.
For the system sizes reported in Fig. \ref{fig6}, $s(t)$ converges to an asymptotic slope $\tilde s$	on a timescale, which 
is order $N$ for both kinds of initial condition. For small pinning frequencies $\nu^2=0.1$ (panel (a)), $\tilde s$ is tiny,
as one can imagine that the scattering of Toda solitons induced by the pinning potential is very weak. For larger $\nu$ 
(panels (b) and (c)), $\tilde s$ approaches larger and larger values upon increasing the system size $N$, although the value
 expected for Fourier profiles, $\tilde s=(T_R-T_L)$  is never observed with these parameters. 

Altogether, we can conclude that heat transport in the pinned Toda chain is affected by very important finite size effects.
For relatively large chains and a rather broad range of pinning strengths, energy transport through the chain is
dominated by a ballistic contribution, due to
traveling Toda-like solitons, that are very weakly affected by the
soft pinning potential. The propagation speed of such structures is quite larger than the speed of acoustic waves
in the lattice, they are therefore allowed to travel very large distances before being ``scattered" by the
pinning component. Since in particularly short chains the scattering effect is practically ineffective, the Toda-like
solitons provide the major contribution to a quasi ballistic hydrodynamic behavior.\\
\section{Summary and perspectives}
The first conclusion of the work is that the Toda chain with harmonic pinning
is no exception to the belief that chains with on-site potential belong 
to the class of normal (Fourier) heat diffusion. From this point of
view the additional integral of motion (\ref{conserv}) does not seem 
to play a crucial role. 
Although the model
is chaotic in the usual sense (Section \ref{sec:nonint}), it displays however some form of 
solitonic transport, yielding a relatively slow and unexpected decay
of current correlations. In this respect, we should also mention
the case of the Fermi-Pasta-Ulam-$\beta$ model 
equipped with a substrate quadratic potential that unexpectedly exhibits  
landmarks of heat superdiffusion \cite{Xiong2018}. 
Actually, it should be noted that the presence of the underlying Toda
potential seems essential here. This phenomenology calls for a deeper
theoretical explanation, to understand why the quadratic pinning 
is so special and why coherent transport is so weakly
affected by it.

Another conclusion regards the general question of the effects of 
perturbations on an otherwise integrable system. We have shown in
Section \ref{sec:todacur} that the Toda model nicely fits into
the class of interacting integrable systems but also that different
types of modifications of its Hamiltonian can yield very different results
and qualitatively change the decay laws of current correlations
(Section \ref{sec:pincur}).
We admit that we have no general arguments to account for those 
observations, that deserve further study. 

Finally, another open problem concerns the issue of finite-size effects. 
After long practice of the sport of numerical simulations, both 
at and out of equilibrium, one reaches the conclusion that finite size and finite time effects are quite difficult to be properly controlled in the problem of heat conduction and should be handled with care (see Section \ref{sec:noneq}). 
On the other hand, when dealing with relatively small systems, like nanowires, nanotubes, single molecules,
graphene layers and similar nanosystems, finite size effects become a 
problem of intrinsic physical interest. For both of these reasons, it would be highly desirable having
at disposal a theoretical approach that accounts for sub-leading terms, beyond asymptotic predictions concerning long-time behavior of very large, i.e.
truly macroscopic, systems. This said, it should be pointed out that this is quite a
difficult task to be accomplished, if one considers the many difficulties already
encountered for producing an asymptotic hydrodynamic approach. In order to take into
account finite size effects one should be able to build up a ``second-order" 
hydrodynamic theory out of
a first-order one, that already revealed quite difficult to be properly worked out.
\section*{Acknowledgements}
SL acknowledges a useful discussion and correspondence with H. Spohn.
\bibliography{biblio2}
\end{document}